\documentclass[12pt,preprint]{aastex}

\shorttitle{ULX M82 X-1}
\shortauthors{Okajima et al.}

\begin{document}

\title{A Stellar-mass  Black Hole 
in the Ultra-luminous X-ray Source M82 X-1?}

\author{Takashi Okajima\altaffilmark{1}}
\affil{Code 662, Astrophysics Science Division, NASA's Goddard Space Flight Center, Greenbelt, MD 20771}

\author{Ken Ebisawa}
\affil{Institute of Space and Astronautical Science, 3-1-1 Yoshinodai, Sagamihara, Kanagawa, 229-8510, 
Japan}

\and

\author{Toshihiro Kawaguchi}
\affil{Department of Physics and Mathematics, Aoyama Gakuin University, 
Fuchinobe 5-10-1, Sagamihara, Kanagawa 229-8558, Japan}

\altaffiltext{1}{Department of Physics and Astronomy, The Johns Hopkins University}

\begin{abstract}

We have analyzed the archival {\em XMM-Newton}\/ data of the
bright Ultra-Luminous X-ray Source (ULX) M82 X-1 with an 105 ksec
exposure when the source was in the steady state.  Thanks to the high photon
statistics from the large effective area and long exposure, 
we were able to discriminate  different X-ray
continuum spectral models.  Neither the standard
accretion disk model (where the radial dependency of the disk effective
temperature is $T(r) \propto r^{-3/4}$) nor a power-law model gives
a satisfactory fit. In fact, observed curvature of the M82 X-1 spectrum
was just between those of the two models.  When the exponent of the radial
dependence ($p$ in $T(r) \propto r^{-p}$) of the disk temperature
is allowed to be free, we
obtained $p =0.61^{+0.03}_{-0.02}$.  Such a reduction of $p$ from the
standard value 3/4 under extremely high mass accretion rates is
predicted from the accretion disk theory as a consequence of the
radial energy advection.  Thus, the accretion disk in M82 X-1 is
considered to be in the {\em Slim disk} state, where an optically thick
Advection Dominant Accretion Flow (ADAF) is taking place.  We have
applied a theoretical slim disk spectral model to M82 X-1, and
estimated  the black hole mass $\approx 19-32 M_\odot$.
We propose that M82 X-1 is a
relatively massive stellar black hole which has been
produced through  evolution of an extremely massive star, shining at 
a super-Eddington luminosity by several times the Eddington limit.

\end{abstract}


\section{Introduction}

Ultra-luminous X-ray Sources (ULXs) in nearby galaxies have typical
X-ray luminosities from $10^{39}$ to $10^{41}$ erg s$^{-1}$ (e.g.,
Makishima et al. 2000; Ptak \& Colbert et al. 2004).  M82 X-1 is the
most luminous ULX which is located off the nucleus of the galaxy and has
exhibited X-ray flares as bright as $\sim 10^{41}$ erg s$^{-1}$
(Matsumoto \& Tsuru 1999; Matsumoto et al.\ 2001; Kaaret et
al.\ 2001).  If one assumes that its luminosity is less than the
Eddington luminosity ($L_{\rm Edd}$), the mass of the central object
must be at least $\sim 700 M_\odot$.  Hence, M82 X-1 has been
considered an {\it intermediate}\/ mass black hole candidate
(Matsumoto et al.\ 1999; Kaaret et al.\ 2001).

However, M82 X-1 exhibits  X-ray energy spectrum
which is much harder than what is expected from standard accretion
disks around intermediate black holes. In
fact, the characteristic color temperature of the standard disk
shining at the Eddington luminosity is $\approx 1 $ keV $(M/10\;
M_\odot)^{-1/4}$, which has been confirmed through observations of
many Galactic black hole candidates.  
Therefore, if M82 X-1 has the standard disk around an
intermediate mass black hole with  $\gtrsim 700 M_\odot$, the characteristic disk temperature
is expected to be $\lesssim$ 0.3 keV.
To the contrary, M82 X-1 indicates much harder, power-law type
spectrum (e.g., Strohmayer \& Mushotzky 2003; 
Fiorito \& Titarchuk 2004; Agrawal and Misra 2006).
Such apparently high disk temperatures  have been also reported from
other ULXs  (e.g., Okada et al.\ 1998; Makishima et al.\ 2000).

There are two major models to explain the ``too hot a disk'' problem
of M82 X-1 and other ULXs.
  The first model assumes that the accretion disk is not in
the standard disk state where the gravitational energy released is
converted into optically thick radiation, but in a {\em slim disk}\/
state where radial energy advection is dominant (Watarai, Mizuno \&
Mineshige 2001; Mizuno, Kubota \& Makishima 2001; Ebisawa et
al.\ 2003).  The competing model assumes that the ULX disks have low
temperature ($\lesssim 1 $ keV) as expected for intermediate mass
black holes.  Such disks are assumed to be shrouded by hot, Compton
thick clouds, and the observed X-ray spectra above $\sim 1 $ keV are
due to inverse Compton process (approximated by a power-law) of the
seed photons from the low temperature disk (e.g., Miller et al.\ 2003;
Miller, Fabian \& Miller 2004; Fiorito \& Titarchuk 2004; Wang et al.\ 2004).

In general, it is difficult to distinguish the two competing ULX spectral
models, since these two models have similar spectral shapes in the energy
range where most X-ray sensors are sensitive.
In this paper, we present a precise spectral analysis of 
{\em XMM-Newton}\/ data of M82 X-1 for a 105 ksec exposure. Thanks to the much better statistics than previous observations, we were able to tightly constrain the spectral models.
We show the slim disk model can explain the M81 X-1 energy spectrum above
$\sim3 $ keV (where contamination of star-burst component is negligible), 
which suggests presence of a {\em stellar}\/ black hole 
in the center of M82 X-1.

\section{Observation and Data Analysis}

There are three archival {\em XMM-Newton}\/ data sets of M82. Two
observations were made on May 6, 2001; for 10 ksec (ObsID=0112290401)
and 29 ksec (ObsID=0112290201), respectively. The other was made on
April 21, 2004 for 105 ksec (ObsID=0206080101), which we analyze in
the present paper.  The observation was carried out employing the
European Photon Imaging Camera (EPIC) PN and MOS in the full window
and medium filter mode. Data screening, region selection and event
extraction were performed with the standard software package XMM-SAS v
6.1.0. In order to eliminate possible contamination from solar flares,
events were selected only when the total off-source count rate is less
than 0.17 (MOS) and 0.55 (PN) counts s$^{-1}$ in 10 -- 15\,keV. This
leaves 63 (MOS1), 65 (MOS2) and 50 (PN) ksec of useful time with an
average count rate of 0.7 (MOS) and 2.2 (PN) counts s$^{-1}$. In this
paper, we will primarily analyze the PN data, which have better
statistics.  The MOS data gives the same results with slightly larger
statistical errors.

We extracted the spectrum of M82 X-1 within a radius of $18''$ around
the point source; this procedure is the same as described in Fiorito
et al. (2004) and Strohmayer et al. (2003).  {\it XMM-Newton}'s
moderate spatial resolution ($13''$--15$''$ in half power diameter)
does not allow us to fully resolve the surrounding faint sources
resolved by {\it Chandra}, including sources 4, 5, and 6 in Matsumoto
et al.\ (2001).  However, sources 4 and 6 are always at least a factor
of 10 fainter than M82 X-1, and source 5 was a factor of 3.4 fainter
when M82 X-1 was the faintest (Strohmayer et al. 2003).  Thus,
contamination from these point sources to our M82 X-1 spectral
analysis is insignificant.  {\it Chandra} also revealed diffuse
emission in the central region of M82, which is obvious in the EPIC
images and spectra as well. 
In order to concentrate on the point source spectral analysis, 
we limit our spectral fitting to the energy range $E >$ 3 keV,
where we estimate the diffuse flux less than 10 \% of the point source flux.
The background spectrum was extracted from a ring of the $2'$ outer
radius and the $18''$ inner radius (within which the M82 X-1 spectrum
was extracted), and it was subtracted from the source spectrum after
being normalized to the detector area.  

The pulse-height spectral data were binned by 32 channels, which
correspond to twice the energy resolution (FWHM).  Fittings were
performed from 3 to 11 keV using xspec 
v.11.3.2. We did not include the interstellar absorption model, since
including of which does not affect the fitting result above 3 keV at all.
First, we employed a power-law model and found that there is a weak iron emission line near 6 keV which
may be modeled by a single Gaussian.
 The photon index is found to be 1.73. The Gaussian line is
centered at 6.61\,keV, and has the equivalent width 87\,eV. 
Such an iron emission line may originate either
from  a disk reflection or a diffuse star-burst component, but an investigation
for its origin is beyond the scope of current paper.
We find
$\chi^2$ to be 96 with 43 d.o.f. Next, we applied the disk blackbody
model (Mitsuda et al.\ 1984) to approximate a standard accretion disk
(Shakura \& Sunyaev 1973), including a Gaussian line with similar
parameters as  above. The disk blackbody temperature is found to
be $2.77\pm0.07$\,keV (90\% confidence level for a single parameter hereafter), and $\chi^2 =
81$ (43 d.o.f.).  Other fitting parameters are shown in Table
\ref{fit_table}.

Both the power-low model and the disk blackbody model are rejected
with a confidence level of 99.98\%.  Importantly, if we compare the
two model fits, we notice opposite trends in the residuals
(Fig.\ 1). Namely, the observed spectrum is slightly ``curved''
downward while the power-law, of course,  does not. Also, the observed curvature
is not as large as that of the disk blackbody model.  This indicates
that the observed spectral curvature is just between that of the
power-law model and the disk blackbody model.

Therefore, we then attempted the ``{\it p-free}'' disk model (Mineshige et
al.\ 1994; Hirano et al.\ 1995; Kubota and Makishima 2004; Kubota et
al.\ 2006), where the temperature profile of the accretion disk is given
as $
T (r) = T_{in} \left(r/r_{in}\right)^{-p}
$
with $r_{in}$, $T_{in}$, and $p$ being free parameters\footnote{This model
is now available in the standard xspec v.12.3.0 or later with the name
``diskpbb''.}. The disk
blackbody model has $p=0.75$, and a smaller $p$ value reduces the
spectral curvature and make the spectral shape closer to the
power-law.  We found the best fit parameters $p=0.61^{+0.03}_{-0.02}$,
$T_{in}=3.73^{+0.58}_{-0.40}$\, keV with $\chi^2 = 55$ (42
d.o.f.). The Gaussian parameters are almost the same as those of the
other two models. Statistically, the {\it p}-free disk model describes
the spectral shape best among the three models.  We can calculate the
$F$-value as a measure of the improvement of $p$-free model relative
to the disk blackbody model. We find $F(1,42)=\Delta
\chi^2/\chi^2_\nu= (81-55)/(55/42) = 19.9$. Thus, the improvement of
the $p$-free model over the disk blackbody model is significant with the
99.99 \% confidence.

When the disk luminosity is as high as the Eddington luminosity, an
optically-thick Advection Dominated Accretion Flow (ADAF) appears
(Abramowicz et al.\ 1988).  Such a flow, often called a {\it slim
  disk}, has very low radiation production efficiency due to
photon trapping (Begelman 1978). The low energy spectrum of the slim disk
has the form $L_E \propto E^{-1}$ (Fukue 2000), while that of
the standard disk is $L_E \propto E^{0.33}$.  Since the disk spectral
shape is related to the radial exponent $p$ as $L_E \propto
E^{3-2/p}$, the spectral change from the standard disk to the slim
disk is equivalent to the reduction of $p$ from 0.75 to 0.5 (Watarai
et al. 2000).

Our result of $p=0.61^{+0.03}_{-0.02}$ strongly suggests that energy
advection is actually taking place, and that the M82 X-1 disk is not
in the standard state, but in a slim disk state.  In this paper, we
employ our own slim disk model\footnote{This model is available at
  http://heasarc.gsfc.nasa.gov/docs/xanadu/xspec/models/slimdisk.html
  for use in xspec} (Kawaguchi 2003) to study M82 X-1
energy spectrum.  Kawaguchi (2003) has calculated slim disk spectra
under four different assumptions.  In Model 1, the local emission is
assumed to be modified blackbody, and in Model 2, Comptonization is
taken into account. Gravitational redshift is included in Model 3; and
in Model 4, which is our ``best'' model, transverse Doppler effects
are additionally considered.  In Figure 2, we compare the simulated
Model 4 spectrum with power-law, disk blackbody, and $p$-free disk
model. It is obvious that the simulated Model 4 spectral shape is
well-represented by the $p$-free model, and its curvature is just
between those of the power-law and disk blackbody.  Comparing Figure 1
and 2, we can see that M82 X-1 and the simulated Model 4 share similar
spectral characteristics.

Next, we directly fit the M82 X-1 spectrum with our slim disk model.
We try all four models with assumptions as listed in Table 1. We tried
models with the viscous parameter $\alpha=$ 0.001, 0.01, 0.1 and 1,
and found that $\alpha=1$ gives the best-fit.  We also fit
allowing $\alpha$ to be free, with little improvement of the
fit. Thus, we show only the results with $\alpha = 1$ in
Table 1.
Fixing the distance to M82 at 2.7 Mpc (e.g., Rieke et al. 1980), there
are then only two free parameters, $M$ and $\dot M$. Kawaguchi's model
calculates the face-on disk flux, so we assume the face-on geometry in
the following.

As summarized in table 1, we obtain $M = (19 - 32) \,M_\odot$,
$\dot{M} = (320 - 560) \times L_{\rm Edd}/c^2$
depending on the physical processes assumed.  In the case of the
standard optically thick accretion disk where the inner disk radius is
three times the Schwartschild radius, $\dot{M} = 17.5 \; L_{\rm
  Edd}/c^2$ gives the Eddington luminosity; so we can see that
M82 X-1 has extremely high mass accretion rates. However, since
slim disks are radiation inefficient, the disk luminosity is not so
large as being proportional to the mass accretion rates.  Bolometric
face-on flux $f_{bol}$ is obtained as $\sim 3 \times 10^{-11}$ erg
s$^{-1}$ cm$^{-2}$ by numerically integrating the best-fit model
spectra over the energy, which is weakly dependent on the assumptions.
The bolometric disk luminosity is $L_{bol}=2 \: \pi \: d^2\: f_{bol}$
where $d$ is the distance (2.8 Mpc), and we obtain $L_{bol} \approx 1.4 \times
10^{40}$ erg s$^{-1}$.  Hence, depending on the assumptions, our slim
disk model fits suggest that M82 X-1 is shining at 4 to 6 times the
super-Eddington luminosity.


\section{Discussion}

 We have studied the M82 X-1 spectrum using archival {\em XMM-Newton}
 data of 105 ksec exposure.
We
 have applied the slim disk spectral model of Kawaguchi (2003), and
 estimated the mass $M\approx(19 - 32) \,M_\odot$ and the bolometric luminosity
 4 to 6 times the Eddington luminosity.
Since ULXs are, by definition, very luminous objects, it is rather
straightforward  that their accretion disks are in the slim disk state,
rather than the standard state. While standard accretion disks around
$\gtrsim20 M_\odot$ black holes have characteristic temperatures
$\lesssim$ 0.8 keV (see Section 1), slim disks can explain the
observed high disk temperature ($\sim$2.8 keV).


We briefly review why a slim disk can produce such a hard spectrum
(see Kawaguchi 2003 for more detail).  (1) As the mass accretion rate
increases, the innermost radius of the slim disk can be smaller  than
three times the Schwartschild radius even in the Schwartschild
geometry (Watarai et al.\ 2000), which makes  the innermost disk temperature
higher.  (2) The ratio of the electron scattering opacity to
the absorption opacity increases with mass accretion rates.  Thus,
photons generated deeper in the disk, where the temperature is higher,
can escape from the disk surface more easily, and the local spectral
shape gets closer to  modified blackbody, rather than the
standard blackbody (e.g., Rybicki \& Lightman 1979).  (3) Furthermore,
because of the small absorption, 
inverse Compton scattering (disk Comptonization) is
enhanced to shift energies from electrons to emerging photons.
Above (1) increases the disk effective temperature, while (2) and
(3) increases the spectral hardening factor which is  ratio of 
the local color temperature and the effective temperature.

Based on the slim disk model fitting, we have found M82 X-1 is
shining at 4 to 6 times the Eddington luminosity.  Although the standard
disk cannot exceed the Eddington limit, such a moderate
super-Eddington luminosity is naturally explained in the slim disk
model (e.g., Abramowicz et al. 1988; Watarai et al.\ 2000). Also, a recent
two-dimensional radiation-hydrodynamic numerical simulation reports
that a slim disk is formed under supercritical accretion flow, and
the disk luminosity can exceed the Eddington luminosity by several
factors (Ohsuga al.\ 2005). 

We have estimated the black hole mass in M82 X-1 as $\approx 19-32 M_\odot$.
Although we have not seen such a rather heavy stellar-mass black hole
in our Galaxy, such black holes are not prohibited by stellar
evolution theory (e.g., Fryer 1999). Actually, a universal
luminosity function for X-ray binaries extends toward the highest
luminosity $\sim 10^{40}$ erg s$^{-1}$ without any break (Grimm,
Gilfanov \& Sunyaev 2003), and it is likely that 
ULXs correspond to the
highest luminosity X-ray binaries. This scenario agrees with the
binary evolution synthesis model of ULXs by Rappaport, Podsiadlowski
\& Pfahl (2005), which strongly suggests ULXs are stellar-mass black
hole binaries. Finally, detailed evolution
models of stellar binaries show very small generation rate of
intermediate-mass black holes (Madhusudhan et al. 2006), which is also in favor
of our stellar-mass ULX model.

In conclusion, we 
suggest that the brightest ULX M82 X-1 harbors a rather heavy but
still stellar-mass black hole shining at several times the Eddington
luminosity.  The slim disk model is reasonably successful in
explaining the X-ray energy spectrum of M82 X-1 above $\sim$3 keV.
The high disk
luminosity and temperature, which are characteristics of the slim disk,
 are not specific to M82 X-1, but
also seen from some  other ULXs. We propose that those ULXs having
similar disk properties may also be interpreted in the framework
of the slim disk model with
stellar black holes shining at super-Eddington luminosities.

Finally, we remark that the
present data analysis  of M82 X-1 was limited above
$\sim$ 3 keV in order to avoid
 cotamination from the soft star-burst component, whereas the
disk spectra from putative ``intermediate-mass black holes''
would be most prominent below $\sim$ 3 keV.
Therefore, it will be interesting to apply our slim disk scenario to 
other ULXs in which their disk spectra are clearly  seen  below 
3 keV as well as above 3 keV. 
If the slim disk model 
is sucessful to explain the ULX energy spectra in the entire energy range, 
intermediate-mass black holes are not required  to explain
the X-ray energy spectra of ULXs.

\acknowledgments

TO acknowledges support from NASA Grant NNG04GB78A. TK thanks the
financial supports from the Japan Society for the Promotion of Science
(JSPS) Postdoctoral Fellowships. This research has made use of public
data and software obtained from the {\it XMM-Newton} Science Archive
(XSA), provided by the European Space Agency (ESA), and the High
Energy Astrophysics Science Archive Research Center (HEASARC),
provided by NASA's Goddard Space Flight Center. The authors thank John
P. Lehan for useful comments and assistance in correcting grammatical
errors in the manuscript.

\clearpage

\begin{figure}
 \centerline{\scalebox{0.8}{\plotone{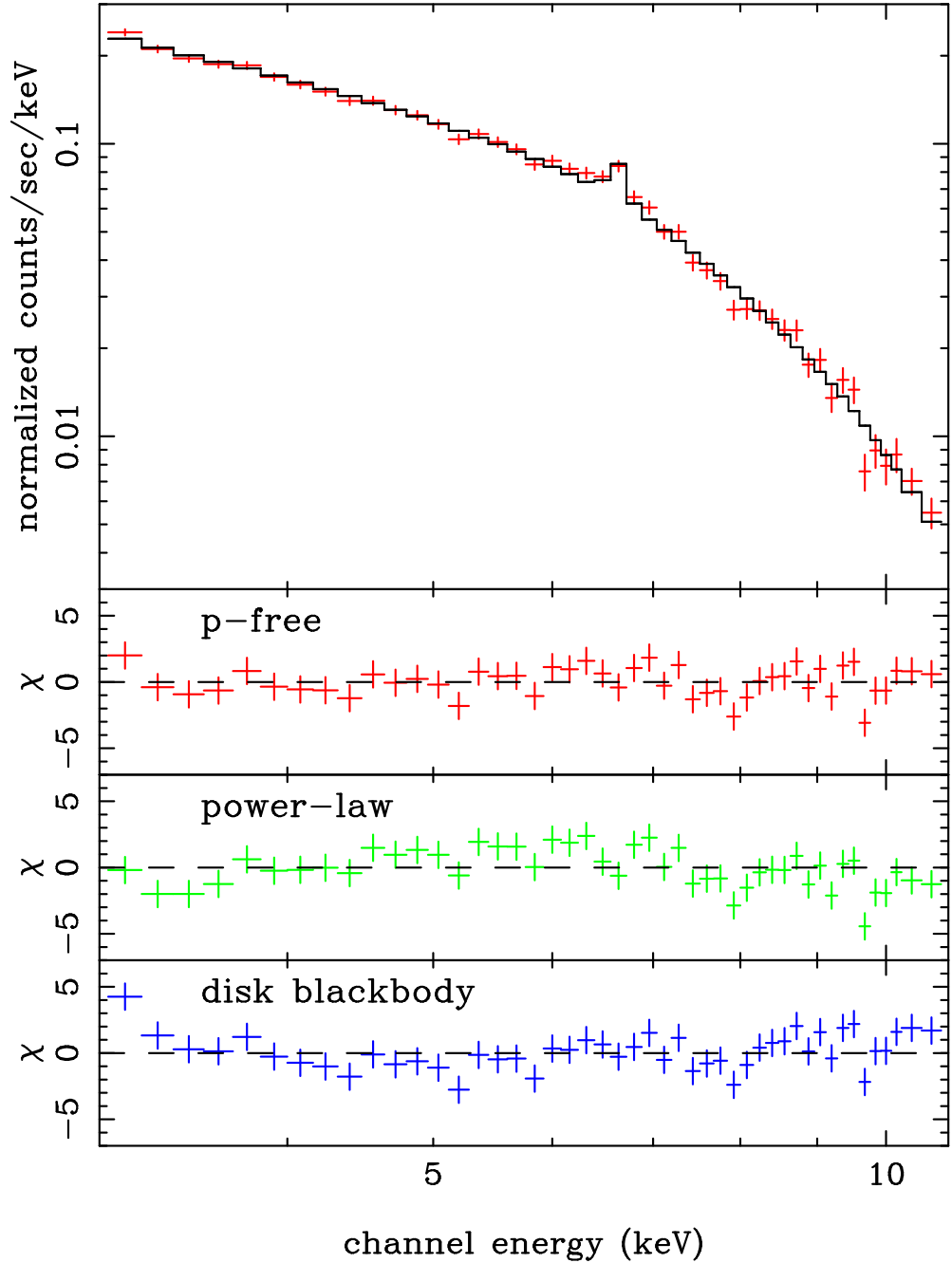}}}
\caption{Folded spectrum of M82 X-1, fitted with the $p$-free disk
  blackbody and narrow Gaussian model (top), and residuals for
  fitting with three different models, $p$-free, power-law, and disk
  blackbody. 
}
\end{figure}

\clearpage

\begin{figure}
 \centerline{\scalebox{0.8}{\plotone{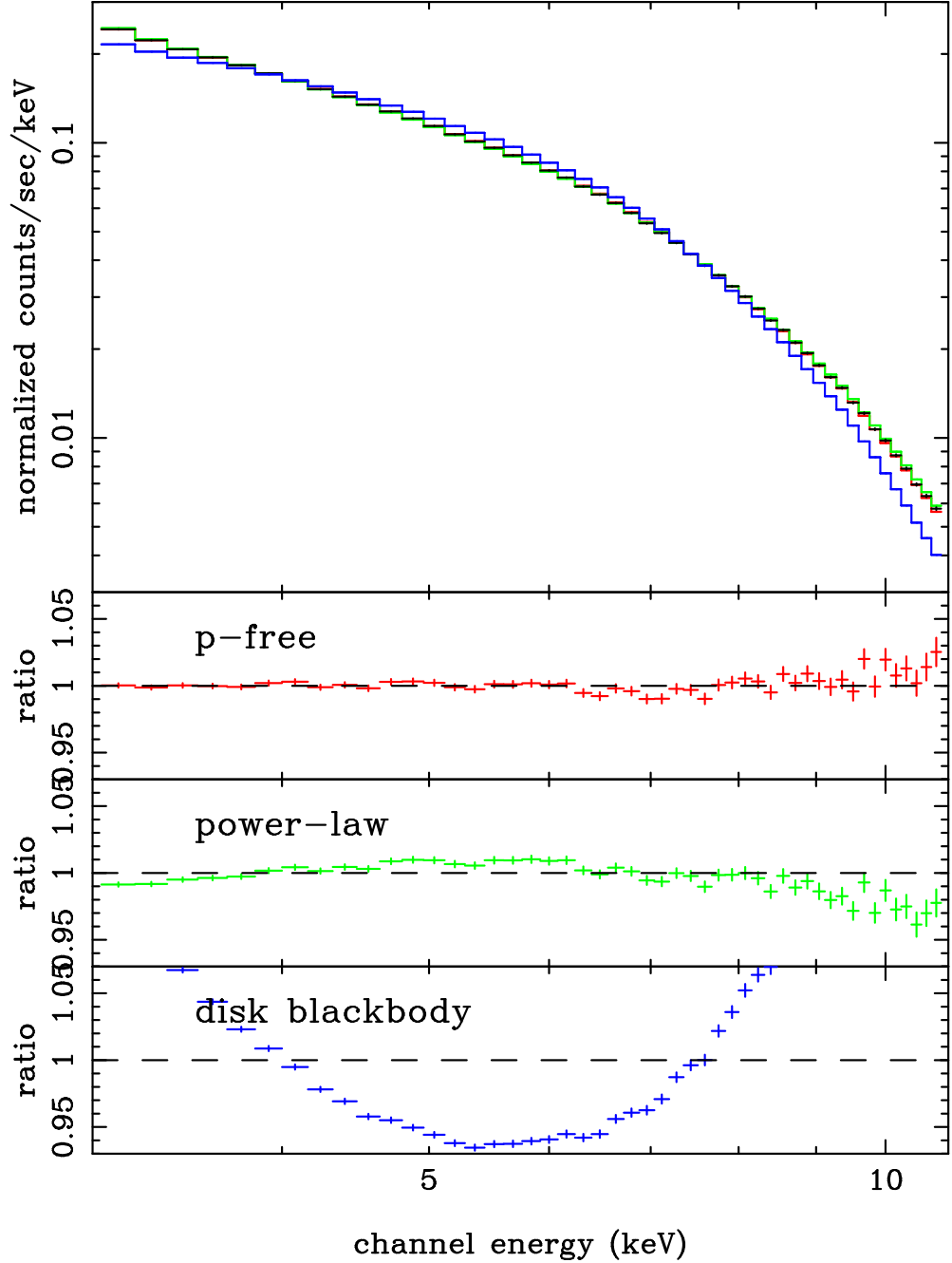}}}
\caption{Comparison of the slim disk model by Kawaguchi (2003)
and other models used to fit M82 X-1.
A simulated spectrum was made from Kawaguchi's model 4 (see text)
with $\alpha=1$, $M=30 M_\odot$ and $\dot M = 350\; L_{Edd}/c^2$ at 2.7 Mpc, and fitted 
with a power-law (index=1.76), disk blackbody ($T_{in}=2.72$), and $p$-free disk 
model ($p=0.54$ and $T_{in}=7.60$).
The top panel shows the simulated slim disk spectrum and the best-fit models,
 and the three bottom panels exhibit the ratios of the simulated data to  the $p$-free model,
power-law and disk blackbody, respectively.
Note that slim disk (solid line) and $p$-free (dashed line) models are almost identical.
}
\end{figure}

\clearpage

\begin{deluxetable}{ccccccccccc}
\tabletypesize{\scriptsize}
\tablecaption{Fitting parameters}
\tablehead{
 & \multicolumn{4}{c}{continuum model parameters} & & &\multicolumn{2}{c}{Gaussian model\tablenotemark{a}} & \\
\cline{2-5} \cline{8-9}
\colhead{model name} & \colhead{$\Gamma$} & \colhead{$T_{in}$} & \colhead{$p$} & \colhead{$N$} & &  & \colhead{$E$} & \colhead{$EW$} &
\colhead{} & \colhead{$\chi^2$/d.o.f.}\\
 & & \colhead{(keV)} & & & & & \colhead{(keV)} & \colhead{(eV)} & &}
\startdata
power-law     & $1.73$ & \nodata & \nodata &  $ 0.00292$\tablenotemark{b} & & & $6.61$ & $87.1$ & & 96/43\tablenotemark{*} \\
disk blackbody & \nodata & $2.77\pm0.07$ & 0.75 & 0.0130\tablenotemark{c}  & & & $6.63^{+0.06}_{-0.05}$ &$61^{+23}_{-18}$ & & 81/43 \\
$p$-free       & \nodata & $3.73^{+0.58}_{-0.40}$\,keV &$0.61^{+0.03}_{-0.02}$ & 0.0028$^{+0.0017}_{-0.0014}$\tablenotemark{c}& & & $6.62^{+0.06}_{-0.04}$ &
$72^{+23}_{-20}$ & & 55/42 \\ 
\\
slim disk ($\alpha=1$)\tablenotemark{d} & \multicolumn{3}{c}{Local Spectral Assumption} & \colhead{$M/M_\odot$}
&\colhead{$\dot{M}/(L_{\rm Edd}/c^2)$} & & & & & \\
\cline{2-6}
1 & \multicolumn{3}{l}{modified B.B.\tablenotemark{e}} & 19 &
395 & & 6.61 & 85 & & 91/43\tablenotemark{*} \\
2 & \multicolumn{3}{l}{Comptonization\tablenotemark{f}} & 19 & 559 & &
6.61 & 85 & & 88/43\tablenotemark{*} \\
3 & \multicolumn{3}{l}{Comptonization$+$gravitational
  redshift\tablenotemark{g}} & $27^{+9}_{-4}$ & $366^{+100}_{-190}$ & &
$6.61^{+0.07}_{-0.03}$ & $86^{+11}_{-22}$ & & 85/43 \\
4 & \multicolumn{3}{l}{Comptonization$+$relativistic
  effects\tablenotemark{h}} & $32^{+6}_{-5}$ & $320^{+60}_{-140}$ & &
$6.61^{+0.07}_{-0.03}$ & $86\pm21$ & & 84/43 \\
\enddata
\tablecomments{Errors correspond to the single parameter  90 \% confidence.}
\tablenotetext{a}{Intrinsic line-width is fixed to 0.01\,keV}
\tablenotetext{b}{Normalization at 1 keV (photons s$^{-1}$ keV$^{-1}$ cm$^{-2}$).}
\tablenotetext{c}{$\left((R_{in}/1\; {\rm km})/(d/10\; {\rm kpc})\right)^2 \cos \theta $ where $d$ is the distance,  $R_{in}$ is the innermost disk radius and $\theta$ is the  inclination.}
\tablenotetext{d}{Source distance is fixed at 2.7 Mpc. }
\tablenotetext{e}{\protect Local emission at each radius is computed considering  electron scattering opacity. See \S 3.}
\tablenotetext{f}{\protect Local emission at each radius is computed considering Compton effects.}
\tablenotetext{g}{Gravitational redshift is included.}
\tablenotetext{h}{In addition to gravitational redshift, the transverse Doppler effect is included.}
\tablenotetext{*}{Errors are not derived because reduced $\chi^2>$2.}
\label{fit_table}
\end{deluxetable}

\end{document}